\documentclass[prd,aps,floats,preprintnumbers,preprint]{revtex4}
\usepackage{graphicx,color}
\usepackage[english]{babel} % grammar package
\usepackage{braket}%Dirac notation
\newcommand{\be}{\begin{equation}}
\newcommand{\ee}{\end{equation}}
\newcommand{\bea}{\begin{eqnarray}}
\newcommand{\eea}{\end{eqnarray}}

\newcommand{\lcdm}{$\Lambda$CDM}
\newcommand{\logA}{\ensuremath{\ln(10^{10}A_s)}}

\usepackage{morefloats} %When there are many Figs.~
\usepackage[titletoc,title]{appendix}
\usepackage{color}
\begin{document}

\title{CMB anomalies and the effects of local  features of the inflaton potential}

\author{Alexander Gallego Cadavid$^{1,5,6}$, Antonio Enea Romano$^{1,2,6}$, Stefano Gariazzo${}^{2,3,4}$}
\affiliation
{
${}^{1}$Yukawa Institute for Theoretical Physics, Kyoto University, Japan\\
${}^{2}$Department of Physics, University of Torino, Via P. Giuria 1, I--10125 Torino, Italy\\
${}^{3}$INFN, Sezione di Torino, Via P. Giuria 1, I--10125 Torino, Italy\\
${}^{4}$Instituto de F\'{\i}sica Corpuscular (CSIC-Universitat de Val\`{e}ncia), Paterna (Valencia), Spain \\
${}^{5}$ICRANet, Piazza della Repubblica 10, I--65122 Pescara \\
%${}^{6}$King's College London, Strand, London, WC2R 2LS, United Kingdom\\
${}^{6}$Instituto de Fisica, Universidad de Antioquia, A.A.1226, Medellin, Colombia
}

\begin{abstract}
Recent analysis of the WMAP and Planck data have shown the presence of a dip and a bump in the spectrum of primordial perturbations  at the scales $k=0.002$ Mpc${}^{-1}$ and $k=0.0035$ Mpc${}^{-1}$ respectively. 
We analyze for the first time the effects a local feature in the inflaton potential to explain the observed deviations from scale invariance in the primordial spectrum. We  perform a best fit analysis of the cosmic microwave background (CMB) radiation temperature and polarization data. The effects of the features can improve the agreement with observational data respect to the featureless model. The best fit local feature affects the primordial curvature spectrum mainly in the region of the bump, leaving the spectrum unaffected on other scales.
\end{abstract}

\maketitle

\section{Introduction}

\iffalse 
The recently released Planck data, however, reported
deviations from a power law connected to a deficit at
multipoles $l \approx 20-40$ in the temperature power spectrum.
\fi
There are important observational motivations to study  modifications of the inflaton potential, like the observed deviations of the spectrum of primordial curvature perturbations from a power law spectrum 
~\cite{Gariazzo:2014dla,Hazra:2014jwa,Hunt:2013bha,Mukherjee:2003cz,Mukherjee:2003ag,Bridle:2003sa,Hazra:2013nca,Vazquez:2013dva, 
Hannestad:2000pm,
Bridges:2008ta,Hazra:2014aea, Gauthier:2012aq,dePutter:2014hza,Ade:2015lrj,Planck:2013jfk,DiValentino:2016ikp,Benetti:2016tvm,Chen:2016vvw,Xu:2016kwz,Chen:2016zuu,Hazra:2016fkm}.
In Refs.~\cite{Gariazzo:2014dla,Hazra:2014jwa,Hunt:2013bha,Mukherjee:2003cz,Mukherjee:2003ag,Bridle:2003sa,Hazra:2013nca,Vazquez:2013dva, 
Hannestad:2000pm,
Bridges:2008ta,Hazra:2014aea, Gauthier:2012aq,dePutter:2014hza} the authors study the effects of analizing the cosmic microwave 
background (CMB) radiation using a free function for the spectrum of primordial scalar perturbations, i.e., they do not consider the usual power law 
spectrum predicted by most of the simplest inflationary models \cite{encyclopaedia,Ade:2015lrj,Planck:2013jfk}. For example, the primordial spectrum 
can be parametrized with wavelets \cite{Mukherjee:2003cz,Mukherjee:2003ag}, linear interpolation \cite{Bridle:2003sa,Hazra:2013nca,Vazquez:2013dva}, interpolating
spline functions \cite{Hazra:2014aea,Gauthier:2012aq,dePutter:2014hza}, among other methods \cite{Gariazzo:2014dla,DiValentino:2016ikp}.

Some interesting evidence of these deviations were given in~\cite{Gariazzo:2014dla,DiValentino:2016ikp} where it was used a method based on a “piecewise cubic Hermite 
interpolating polynomial” 
(PCHIP) for the primordial power spectrum. This analysis showed that the spectrum of primordial perturbations can be approximated with a power law in the range 
of values $ 0.007 \mbox{ Mpc}^{-1} <k < 0.2 \mbox{ Mpc}^{-1}$ while in the range $ 0.001 \mbox{ Mpc}^{-1} <k < 0.0035 \mbox{ Mpc}^{-1}$ there are a dip and a bump at $k=0.002$ Mpc${}^{-1}$ and $k=0.0035$ Mpc${}^{-1}$, with a statistical significance of about $2\sigma$ and $1\sigma$, respectively. Similar results were reported in several other analyses
\cite{Shafieloo:2003gf,Nicholson:2009pi,Hazra:2013ugu,Hazra:2014jwa,
Nicholson:2009zj,Hunt:2013bha,Hunt:2015iua,
Goswami:2013uja,Matsumiya:2001xj,Matsumiya:2002tx,
Kogo:2003yb,Kogo:2005qi,Nagata:2008tk,Ade:2015lrj,
Gariazzo:2014dla,DiValentino:2015zta}
using different techniques and both the WMAP~\cite{wmapfmr} and Planck~\cite{pi,Adam:2015rua} measurements.
%The presence of features in the primordial power spectrum outside the causal horizon at the time of the CMB decoupling cannot be reconduced to any physical phenomena that occurs during the late-time evolution. (late time physics can contaminate primordial pertubatiions effects on the cmb photons, so this statement is not true)
%A feature in the initial power spectrum can then be the most natural explanation for the dip and the bump in the CMB spectrum.
In this paper we study how local features of the inflaton potential can model this type of local glitches of the spectrum of primordial curvature perturbations. We also study the effects of these features on the primordial tensorial perturbation spectrum.

Features of the inflaton potential can affect the evolution of primordial curvature perturbations~\cite{GallegoThesis,Cadavid:2015iya, 
Gariazzo:2014dla,Motohashi:2015hpa,et,aer,a1,a2,a3,Adams,Chluba:2015bqa,Chen:2011zf,Palma:2014hra, starobinsky,constraints1,
constraints2,Hazra:2014jka,Hazra:2014goa,Martin:2014kja,Romano:2014kla,Ashoorioon:2006wc,Ashoorioon:2008qr,Cai:2015xla} and consequently  generate a variation in the amplitude of the spectrum and 
bispectrum~\cite{GallegoThesis,Cadavid:2015iya,Motohashi:2015hpa,et,aer,a1,a2,a3,Adams,Romano:2014kla}. This can  provide a 
better fit of the observational data in the regions where the spectrum shows some deviations from a power law~\cite{Motohashi:2015hpa,et,constraints1,constraints2,a1,a2,a3,Adams,Gariazzo:2014dla,Hazra:2014jwa,Hunt:2013bha,Joy:2007na, Joy:2008qd, 
Mortonson:2009qv}.
In this paper we perform a best fit analysis of the CMB radiation temperature and polarization data and we study the effects of a local feature of the inflation potential which affects the primordial curvature spectrum in the region of the bump.

\section{Local features}
We consider a single scalar field minimally coupled to gravity with a standard kinetic term according to the action
\begin{equation}\label{action1}
  S = \int d^4x \sqrt{-g} \left[ \frac{1}{2} M^2_{Pl} R  - \frac{1}{2}  g^{\mu \nu} \partial_\mu \phi \partial_\nu \phi -V(\phi)
\right],
\end{equation}
where $ M_{Pl} = (8\pi G)^{-1/2}$ is the reduced Planck mass and $g_{\mu \nu}$ is the flat $FLRW$ metric. The Friedmann equation and the equation of 
motion of the inflaton are obtained from the 
variation of the action with respect to the metric and the scalar field respectively
\begin{equation}\label{ema}
  H^2 \equiv \left(\frac{\dot a}{a}\right)^2= \frac{1}{3 M^2_{Pl}}\left( \frac{1}{2} \dot \phi^2 + V(\phi) \right),
\end{equation}
\begin{equation}\label{emphi}
  \ddot \phi + 3H\dot \phi + \partial_{\phi}V = 0,
\end{equation}
where $H$ is the Hubble parameter, and dots and $\partial_{\phi}$ denote derivatives with respect to time and scalar field respectively. 
The slow-roll parameters are defined
\bea \label{slowroll}
  \epsilon \equiv -\frac{\dot H}{H^2} \,\,\,\, , \,\,\,\, \eta \equiv \frac{\dot \epsilon}{\epsilon H}.
\eea
We consider a potential energy given by \cite{Cadavid:2015iya}
\bea\label{pot}
V(\phi)&=& V_{0}(\phi) + V_F(\phi) \, , \\
V_F(\phi)&=&\lambda e^{-( \frac{\phi-\phi_0}{\sigma})^{2}} \, , \label{LF}
\eea
where $V_{0}(\phi)$ is the featureless potential and $V_F$ corresponds to a step symmetrically dumped by an even power negative exponential factor. 
In this paper we will consider the case of a quadratic inflaton potential
\bea 
V_0(\phi)=\frac{1}{2} m^2 \phi^2 \,.
\eea 
The tensor-to-scalar ratio for a monomial potential $\phi^n$ is $r \approx 16 n /(4N_e + n)$, where $N_e$ is the number of $e$-folds before the end of inflation \cite{Ade:2015lrj,Planck:2013jfk}. In the case of quadratic inflation $r\approx 0.16$ for $N_e\approx 50$, which is not in good agreement with observational data. Our analysis confirms this when we fit data without the feature.  We will show later that the effects of local features improve the agreement with CMB data but not enough to get a $\chi^2$ as low as the one of other inflationary models with lower values of $r$.

This type of modification of the slow-roll potential is called local feature (LF) \cite{Cadavid:2015iya} which 
differs from the branch feature (BF) \cite{Cadavid:2015iya,Romano:2014kla} since the potential is symmetric with respect to the location of the 
feature and it is only affected in a limited range of the scalar field value. Due to this the spectrum and bispectrum are only modified in a narrow 
range 
of scales, in contrast to the BF in which there are differences in the power spectrum between large and small scale which are absent in the case of LF.
In some cases the step in the spectrum due to a BF can be very small, and the difference between large and small scale effects would not make BF observationally distinguishable from LF.  Nevertheless in general the oscillation patterns produce in the spectrum by a single BF would be different because a single LF can be considered as the combination of two appropriate BF  \cite{Cadavid:2015iya}.

In this paper we use the local type effect of these features to model phenomenologically local glitches of the primordial scalar spectrum 
on the scales $k=0.002\mbox{ Mpc}^{-1}$ and  $k=0.0035\mbox{ Mpc}^{-1}$~\cite{Gariazzo:2014dla}, and to study their effects on the primordial tensor 
spectrum, without affecting  other scales.

\begin{figure}
 \begin{minipage}{.45\textwidth}
  \includegraphics[scale=0.6]{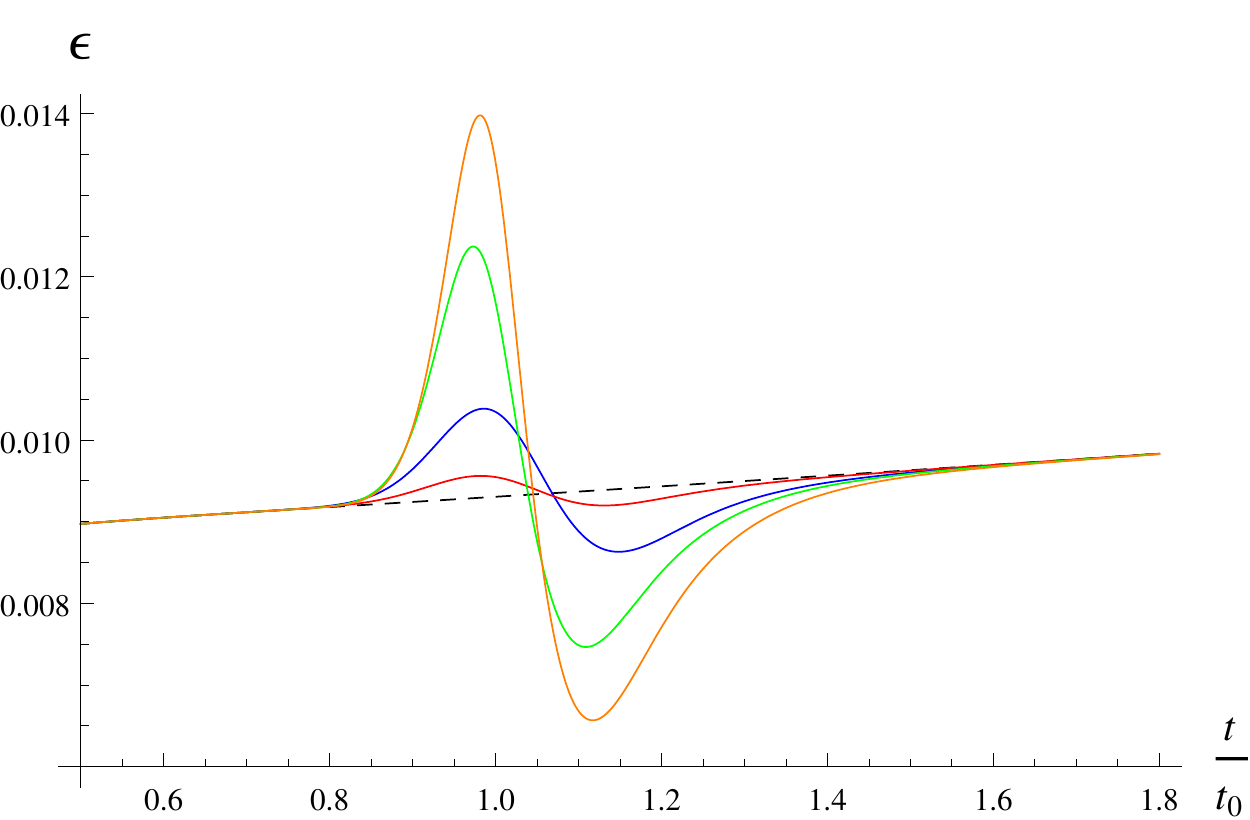}
  \end{minipage}
 \begin{minipage}{.45\textwidth}
  \includegraphics[scale=0.6]{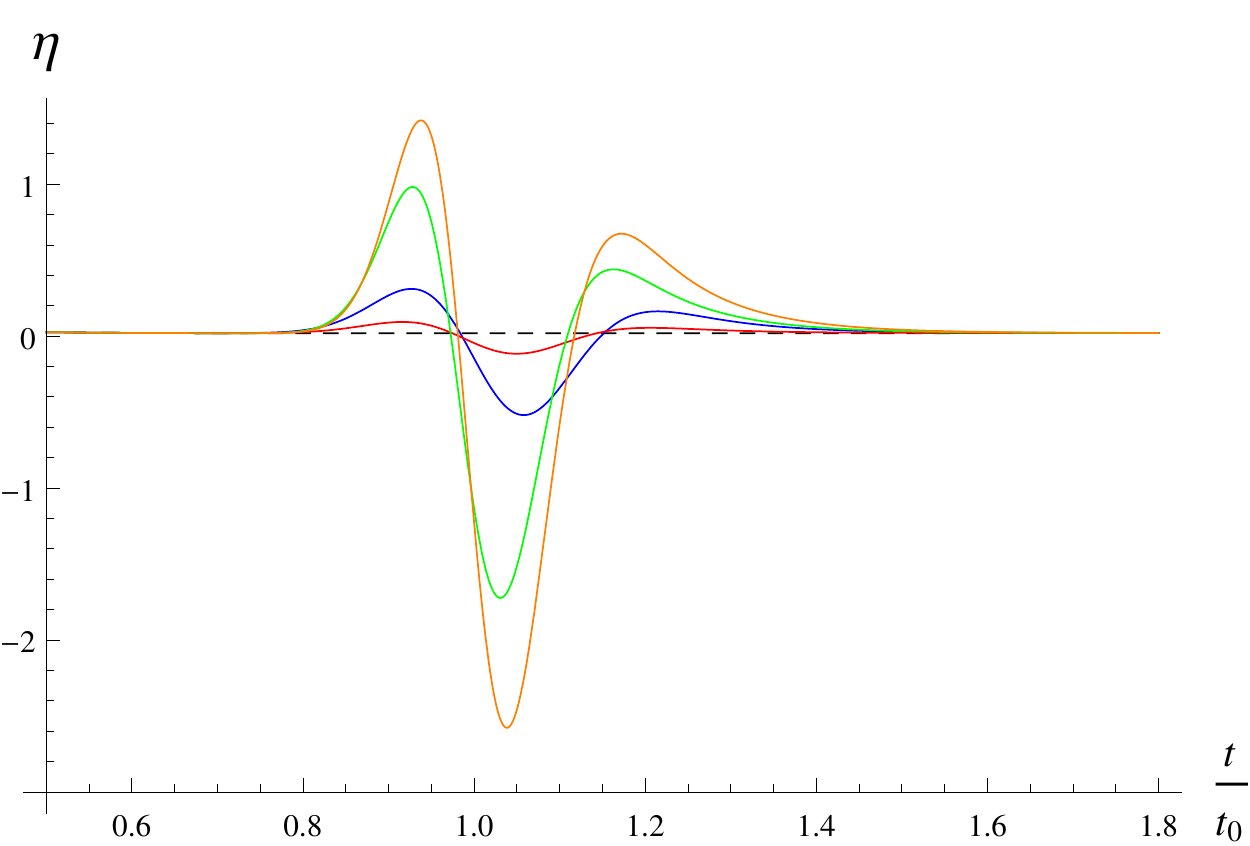}
 \end{minipage}
 \caption{The numerically computed slow-roll parameters $\epsilon$ and $\eta$ around the feature time $t_0$ for $\lambda=-4\times10^{-12}, 
\sigma=0.05$, and $k_0=1.2\times 10^{-3}$ (blue), 
$\lambda=-10^{-12}, \sigma=0.05$, and $k_0=1.3\times 10^{-3}$ (red), $\lambda=-10^{-11}, \sigma=0.04$, and $k_0=1.3\times 10^{-3}$ (green), and  
$\lambda=-1.5\times 10^{-11}, \sigma=0.04$, and $k_0=1.2\times 10^{-3}$ (orange). The dashed lines correspond to the featureless slow roll parameters.
}
\label{slowrollplot}
\end{figure}

The effects of the feature on the slow-roll parameters are shown in Fig. \ref{slowrollplot}, where we can see that there are oscillations of the 
slow-roll parameters around the feature time $t_0$, defined as $\phi_0=\phi(t_0)$ \cite{Cadavid:2015iya}. The magnitude  of the potential modification is controlled by the 
parameter 
$\lambda$, as its effect is such that larger value of $\lambda$ give larger values of the slow-roll parameters. The size of the range of field  values where the potential 
is affected by the feature is determined by the parameter $\sigma$ and the slow-roll parameters are smaller for larger $\sigma$. We define $k_0$ as the scale exiting the horizon at the feature time $t_0$, $k_0=-1/\tau_0$, where $\tau_0$ is the value of conformal time corresponding to $t_0$.  Oscillations occur around $k_0$, and their location can be controlled by changing $\phi_0$. We adopt a system of units in which $c=\hbar=M_{Pl}=1$.

\section{Spectrum of curvature tensor perturbations}\label{scp}
\begin{figure}  \includegraphics[scale=1.2]{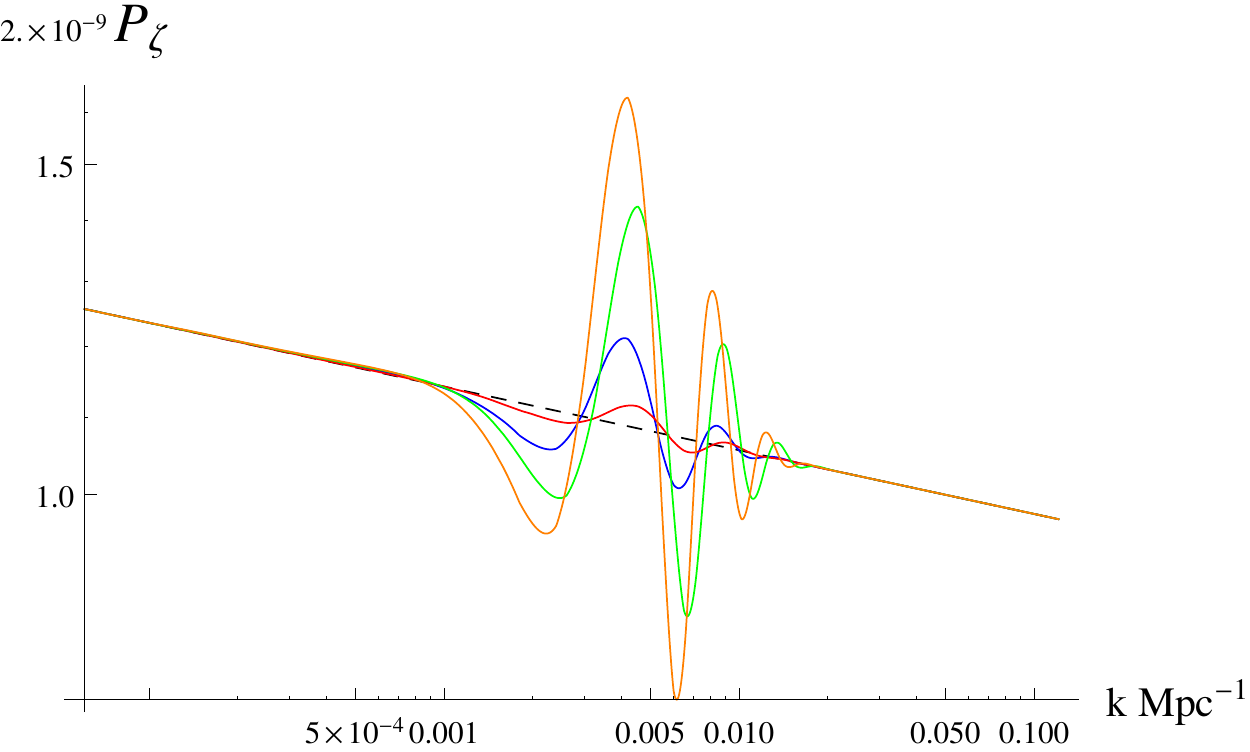}
 \caption{The power spectrum of primordial curvature perturbations $P_{\zeta}$ is plotted for $\lambda=-4\times10^{-12}, \sigma=0.05$, and 
$k_0=1.2\times 10^{-3}$ (blue), 
$\lambda=-10^{-12}, \sigma=0.05$, and $k_0=1.3\times 10^{-3}$ (red), $\lambda=-10^{-11}, \sigma=0.04$, and $k_0=1.3\times 10^{-3}$ (green), and  
$\lambda=-1.5\times 10^{-11}, \sigma=0.04$, and $k_0=1.2\times 10^{-3}$ (orange). The dashed lines correspond to the featureless spectrum.}
\label{Pplot}
\end{figure}

In order to study the curvature perturbations we expand perturbatively the action 
with respect to the background $FLRW$ solution. The second order action for scalar perturbations in the comoving gauge takes the form \cite{m}
\bea
\label{s2}
 S_2&=& \int dt d^3x\left[a^3 \epsilon \dot\zeta^2-a\epsilon(\partial \zeta)^2 \right] \,.
\eea
The equation for  curvature perturbations $\zeta$ obtained from the Lagrange equations is
\begin{equation}
 \frac{\partial}{\partial t}\left(a^3\epsilon \frac{\partial \zeta}{\partial t}\right)- a\epsilon\delta^{ij} \frac{\partial^2 \zeta}{\partial 
x^i\partial x^j}=0.
\end{equation}
Taking the  Fourier transform and using conformal time $d\tau \equiv dt/a$ we get
\begin{equation}\label{cpe}
  \zeta''_k + 2 \frac{z'}{z} \zeta'_k + k^2 \zeta_k = 0,
\end{equation}
where  $k$ is the comoving wave number, $z\equiv a\sqrt{2 \epsilon}$, and primes denote a derivative with respect to the conformal time. The two-point 
function of curvature perturbations is
\begin{equation}
 \Braket{ \hat \zeta(\vec{k}_1, t) \hat \zeta(\vec{k}_2, t) } \equiv (2\pi)^3 \frac{2\pi^2}{k^3} P_{\zeta}(k) \delta^{(3)}(\vec{k}_1+\vec{k}_2) \, ,
\end{equation}
where the power spectrum of curvature perturbations is defined as 
\begin{equation}\label{ps2}
  P_{\zeta}(k) \equiv \frac{k^3}{2\pi^2}|\zeta_k|^2 \,.
\end{equation}
%\subsection{Effects of the features on the primordial scalar spectrum}
The effects of the features on the primordial scalar spectrum are plotted in Fig. \ref{Pplot} for different values of the parameters $\lambda, 
\sigma$, and $k_0$ \cite{Cadavid:2015iya}. The spectrum of primordial curvature perturbations has oscillations around $k_0$, whose amplitude is 
larger 
for larger $\lambda$ since the latter controls the magnitude  of the potential modification. The amplitude of the spectrum oscillations is larger for 
smaller $\sigma$, because in this case the change in the potential is more abrupt and consequently the slow-roll parameters are larger. % The parameter $k_0$ shifts the spectrum to the right or left for higher or lower values of $k_0$ respectively.

\begin{figure}
 \begin{minipage}{.45\textwidth}
  \includegraphics[scale=0.6]{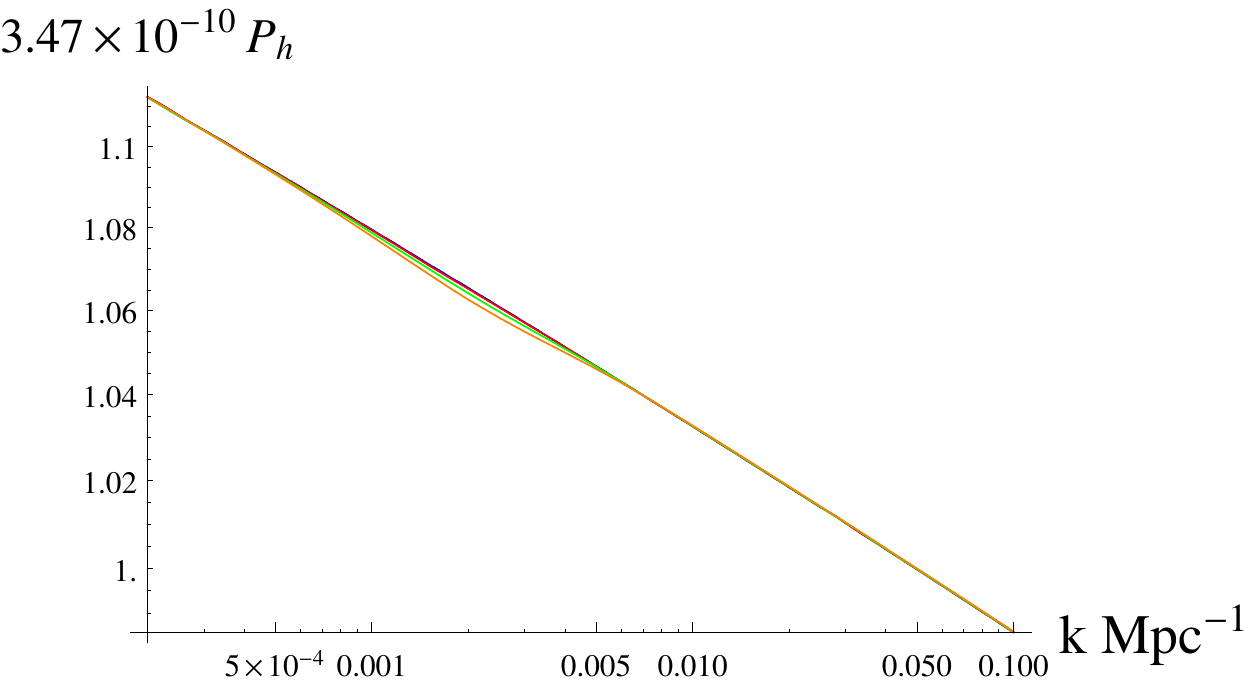}
  \end{minipage}
 \begin{minipage}{.45\textwidth}
  \includegraphics[scale=0.6]{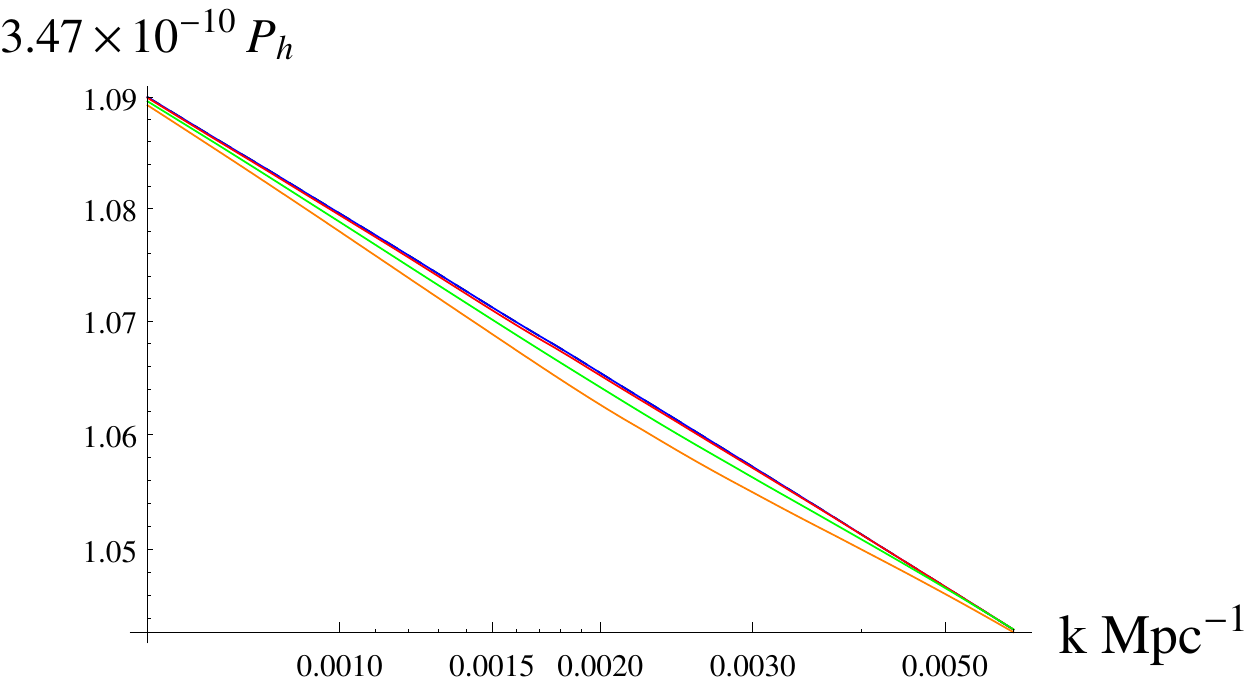}
 \end{minipage}
 \caption{The power spectrum of primordial tensor perturbations $P_{h}$ is plotted for $\lambda=-4\times10^{-12}, \sigma=0.05$, and $k_0=1.2\times 
10^{-3}$ (blue), 
$\lambda=-10^{-12}, \sigma=0.05$, and $k_0=1.3\times 10^{-3}$ (red), $\lambda=-10^{-11}, \sigma=0.04$, and $k_0=1.3\times 10^{-3}$ (green), and  
$\lambda=-1.5\times 10^{-11}, \sigma=0.04$, and $k_0=1.2\times 10^{-3}$ (orange). The dashed lines correspond to the featureless spectrum. The plot on the right corresponds to a zoom of the left plot. As it can be seen the effects of the different features on the spectrum  $P_h$ are rather small and the spectra of the models with features are  difficult to distinguish from the featureless model spectrum.}
\label{Phplot}
\end{figure}

%\subsection{Effects of the features on the CMB spectrum}
%\section{Spectrum of tensor perturbations}
%In this section we study the effects of the feature on the primordial tensor spectrum and the CMB polarization spectrum. Following a similar 
%procedure 
%as in Sect. \ref{scp} we find an equation of motion for the perturbations of the tensor modes $h_k$ given by
The equation for tensor perturbations can be derived in a way similar to the case of scalar perturbations, giving
\bea
h''_k+2\frac{a'}{a} h'_k+k^2 h_k=0 \, ,
\eea
where again $k$ is the comoving wave number. 
The power spectrum of tensor perturbations is obtained from the two-point function as 
\bea
P_{h}(k) \equiv \frac{2k^3}{\pi^2}|h_k|^2\, ,
\eea
from which the tensor-to-scalar ratio can be defined as the ratio between the spectrum of tensor and scalar perturbations as
\bea
r\equiv \frac{P_h}{P_{\zeta}} \, .
\eea
%\subsection{Effects of the features on the primordial tensor spectrum}
The effects of the features on the primordial tensor spectrum are plotted in Fig. \ref{Phplot} for different values of the parameters $\lambda, 
\sigma$, and $k_0$.
These effects are not very significant and in fact the observational data analysis we will present in the rest of the paper confirms that  local features affect mainly the curvature spectrum.
\section{Effects of local features on the CMB spectrum}
\begin{figure}
	\centering
  \includegraphics[width=\textwidth]{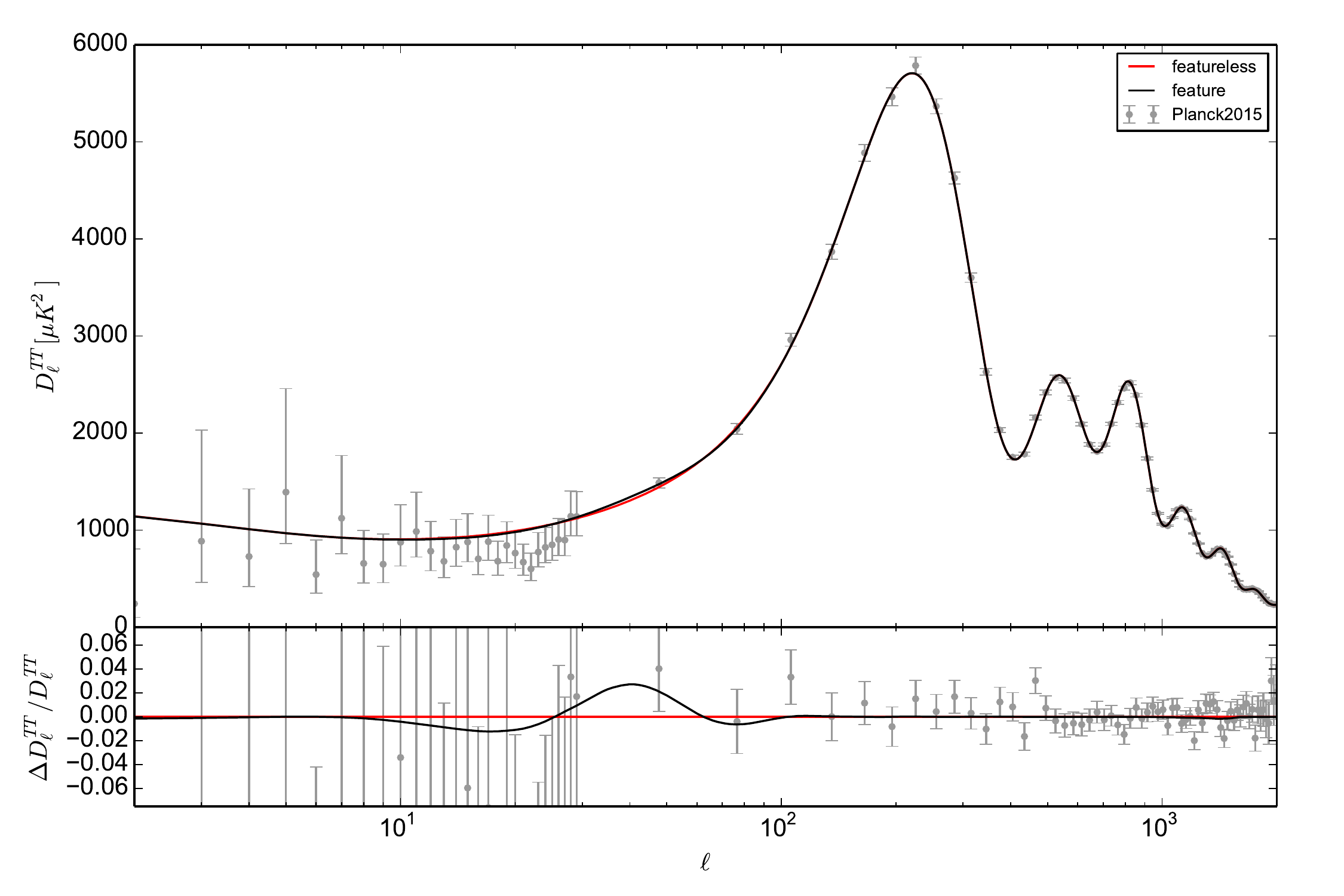}
 \caption{The $D_l^{TT}=\ell(\ell+1) C_\ell^{TT}/(2\pi)$
 spectrum in units of $\mu K^2$ is plotted as a function of the multipole $l$.
 We compare the best-fit obtained using the inflationary model without feature (red line)
 to the one obtained introducing the local feature (black line).
 In the lower panel we plot the relative difference with respect
 to the featureless case.
 The data points are from the 2015 Planck release \cite{Adam:2015rua}.
 The cosmological parameters used to compute the two spectra are reported in Tab.~\ref{tab:results}.
 }
\label{clTT}
\end{figure}

In Fig.~\ref{clTT} %and \ref{clBB}
we show the effects of local features
on the temperature (TT) %and B-mode polarization (BB)
CMB spectrum. %We use the best-fit parameters obtained in the analysis presented in the next sections to compute the theoretical spectra and we compare them with the most recent CMB data.
Since we are considering a feature of local type, as theoretically expected, the spectrum is not affected  on scales sufficiently far from $k_0$.
Branch features \cite{Cadavid:2015iya} could on the contrary also introduce a step in the power spectrum, modifying it also on scales far from $k_0$, and for this reason LF are more appropriate to model local deviations of the spectrum.

The main effects produced by the LF appear between $\ell=10$ and $\ell=100$
in the TT spectrum.
They correspond to the wiggles of the primordial scalar fluctuations  shown in Fig.~\ref{Pplot}.
The class of LF we consider allows to fit the small bump at $\ell\simeq40$
better than the dip at $\ell\simeq20$ in the CMB spectrum.
%Other LF models may also improve the explanation of the dip.
The impact of the LF on the BB spectrum is much smaller,
since, as discussed previously, the effect of the feature on the primordial tensorial perturbations spectrum is negligible.

% \begin{figure}
% 	\centering
%   \includegraphics[width=\textwidth]{clTE.pdf}
%  \caption{As in Fig.~\ref{clTT}, but for the $D_l^{TE}=\ell(\ell+1) C_\ell^{TE}/(2\pi)$ spectrum.}
% \label{clTE}
% \end{figure}
% \begin{figure}
% 	\centering
%   \includegraphics[width=0.9\textwidth]{clEE.pdf}
%  \caption{As in Fig.~\ref{clTT}, but for the $D_l^{EE}=\ell(\ell+1) C_\ell^{EE}/(2\pi)$ spectrum.}
% \label{clEE}
% \end{figure}
% \begin{figure}
% 	\centering
%   \includegraphics[width=0.9\textwidth]{clBB.pdf}
%  \caption{The $D_l^{BB}=\ell(\ell+1) C_\ell^{BB}/(2\pi)$ spectrum is plotted according to the same units and conventions as in Fig.~\ref{clTT}.
%  The purple data points are the lensing B-modes measurements
%  obtained by SPTPol \cite{Keisler:2015hfa}.
% The Bicep2/Keck+Planck (BKP) orange points are obtained in Ref.~\cite{Ade:2015tva} removing from the Bicep2/Keck  spectrum the dust contribution estimated from the Planck 353 GHz maps.
%  }
% \label{clBB}
% \end{figure}

\subsection{The observational data analysis method}
To study the effects produced by local features on the CMB spectrum,
we modified the Boltzmann code \texttt{CAMB}~\cite{Lewis:1999bs}
that computes the theoretical spectra and
the corresponding Markov Chain Monte Carlo (MCMC) code \texttt{CosmoMC}~\cite{Lewis:2002ah}
in order to use a non-standard power spectrum for the primordial curvature perturbations.

As a base model 
we considered the standard parameterization of the \lcdm\ model for the evolution
of the universe, that includes four parameters:
the current energy density of baryons and of Cold Dark Matter (CDM)
 $\Omega_b h^2$ and $\Omega_c h^2$,
the ratio between the sound horizon and
the angular diameter distance at decoupling $\theta$, and the optical depth to reionization $\tau$. 

The parameterization of the primordial power spectra is modified
to take into account the presence of the local feature.
%Since we are considering as our starting point the $\phi^2$ inflationary model,
%the tilt of the primordial power spectrum is fixed ($n_s=0.962$), as well as the tensor-to-scalar ratio ($r=0.158$).
To see the effects of the feature, we compare the results obtained in the
featureless  model
with the ones obtained when a local feature is added.
%We are aware that the most recent CMB data disfavor the $\phi^2$ model
%(see e.g.~Ref.~\cite{Ade:2015lrj}),
%but here we just want to show how a local feature influences the CMB analyses and possibly improve the agreement with observational data.
The comparison of the effects of LF of different inflationary potentials is left for future studies.

The data sets that we use to test the LF
are taken from the last release
from the Planck Collaboration \cite{Adam:2015rua}
for the temperature and E-mode polarization modes.
We consider the temperature and polarization power spectra in the range $2\leq\ell\leq29$ (low-$\ell$) and
only the temperature power spectrum at higher multipoles, $30\leq\ell\leq2500$ (high-$\ell$).
Since the polarization spectra at high multipoles are still under discussion and some residual systematics
were detected by the Planck Collaboration
\cite{Aghanim:2015xee,Ade:2015xua},
we do not include the full polarization spectra obtained by Planck.
Moreover, we do not include the data on the BB spectrum
as obtained from the Bicep2/Keck collaboration~\cite{Ade:2015fwj},
because the baseline inflationary model
that we consider ($\phi^2$) cannot
reproduce the small amount of primordial tensor modes
that are observed after cleaning the Bicep2/Keck data using
the polarized dust emission obtained
by the high frequency maps by Planck~\cite{Ade:2015tva}.
% We call this set of data BKP.
% The full set of data we analyze is the combination of the Planck and BKP data described above.

%As a reference we also give the analysis results obtained \co{assuming a power law spectrum of the type ..} in Ref.~\cite{Ade:2015tva}.
%The differences between the results are driven by the fact that the $\phi^2$
%model predicts a high tensor-to-scalar ratio, that must be compensated
%by a simultaneous variation of all the cosmological parameters.

\section{Results}

\begin{table*}[t]
\begin{center}
\begin{tabular}{lccc}
\hline
Parameter & with feature & featureless \\[0.1cm]
\hline
$10^2\omega_b$ & $2.203\pm0.019$  & $2.203\pm0.019$   \\
$\omega_c$     & $0.122\pm0.001$  & $0.121\pm0.001$   \\
$\theta$       & $1.0406\pm0.0004$& $1.0406\pm0.0004$ \\
$\tau$         & $0.059\pm0.015$  & $0.059\pm0.016$   \\
$\logA$        & $3.056\pm0.031$  & $3.054\pm0.032$   \\
$H_0$          & $66.4\pm0.5$     & $66.5\pm0.5$      \\
\hline
% $n_s$              & $0.962$                 & $0.962$ \\
% $r$                & $0.158$                 & $0.158$ \\
$-10^{12}\lambda$  & $[0.05,1.23)$ $\{1.12\}$& -       \\
$10^2\sigma$       & $5.3\,^{+1.1}_{-3.1}$   & -       \\
$10^3k_0$          & $[1.0,1.3)$ $\{1.13\}$  & -       \\
\hline
$\chi^2_{\text{Planck low-}\ell}$ & 10505.31 & 10504.92\\
$\chi^2_{\text{Planck high-}\ell}$& 764.90   & 767.8  \\
$\chi^2_{\text{nuisance priors}}$ & 1.08    & 0.27    \\
% $\chi^2_{BKP}$                    & 52.15    & 51.91   \\
\hline
$\chi^2_{tot}$    & 11271.29  & 11273.0 \\
\end{tabular}
\end{center}
\caption{Constraints on the cosmological parameters and
$\chi^2$ for the model with and without feature.
All the constraints are given at 1$\sigma$ confidence level.
The lower limits on the feature parameters correspond
to the limits we used as a prior.
The best fit are values inside curly brackets.
We separately report the different contributions to the $\chi^2$
(Planck low-$\ell$, Planck high-$\ell$ %, BKP 
and from the priors on the nuisance parameters) and the total.
%The best fit $\chi^2$ values are typically
%within $\Delta \chi^2 \sim 1$ of the true global best-fit value.
}
\label{tab:results}
\end{table*}
%*********************************
\begin{figure}
	\centering
  \includegraphics[width=\textwidth]{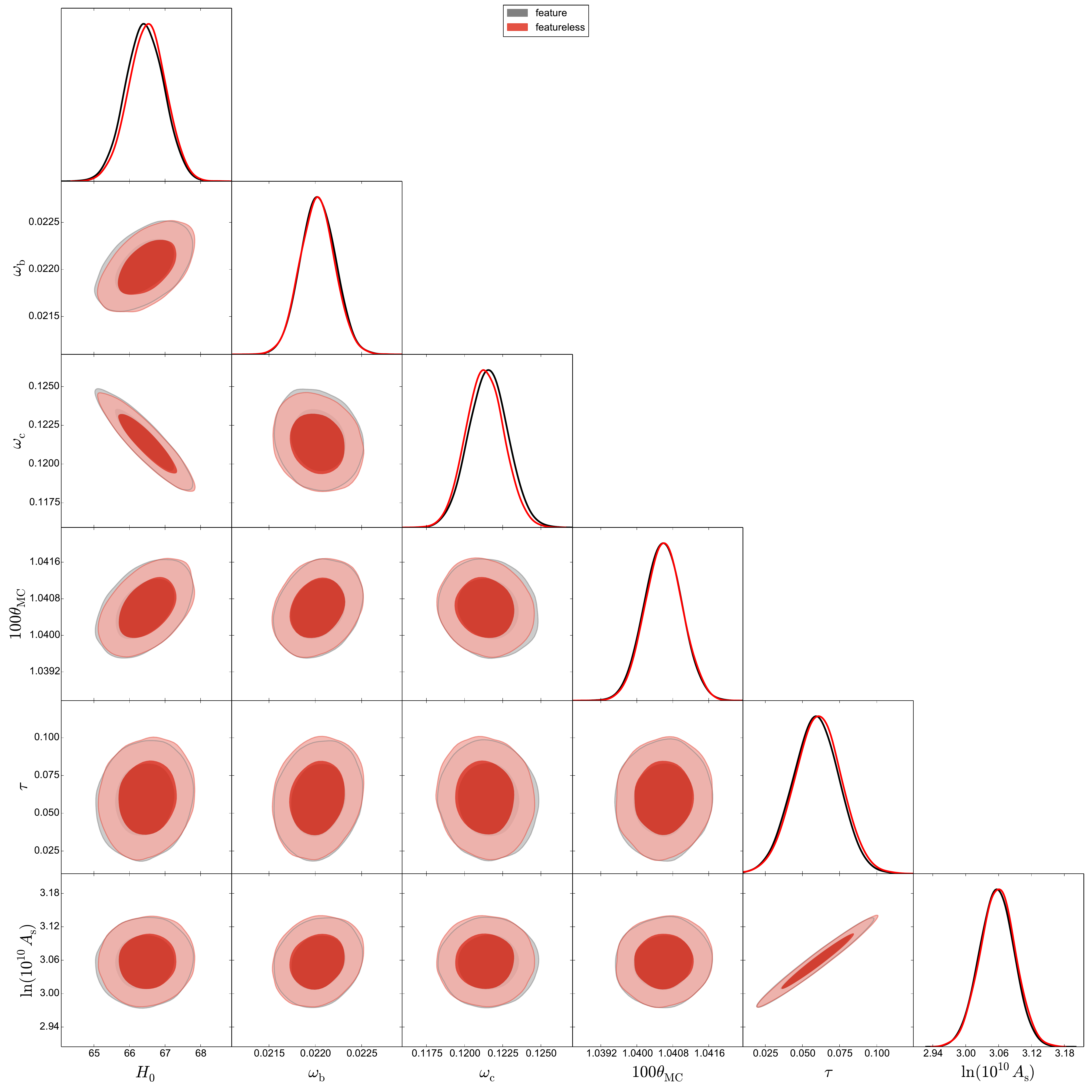}
 \caption{A comparison between the model with and without features is given for the parameters 
 $H_0$, $\omega_b$, $\omega_c$, $\theta$, $\tau$ and $\logA$.
All the results are obtained considering the Planck low-$\ell$+high$\ell$ %+BKP
data combination. As can be seen the effects of the feature on the estimation of these non inflationary cosmological parameters is negligible.
}
\label{fig:lcdm_params}
\end{figure}
\begin{figure}
	\centering
  \includegraphics[width=\textwidth]{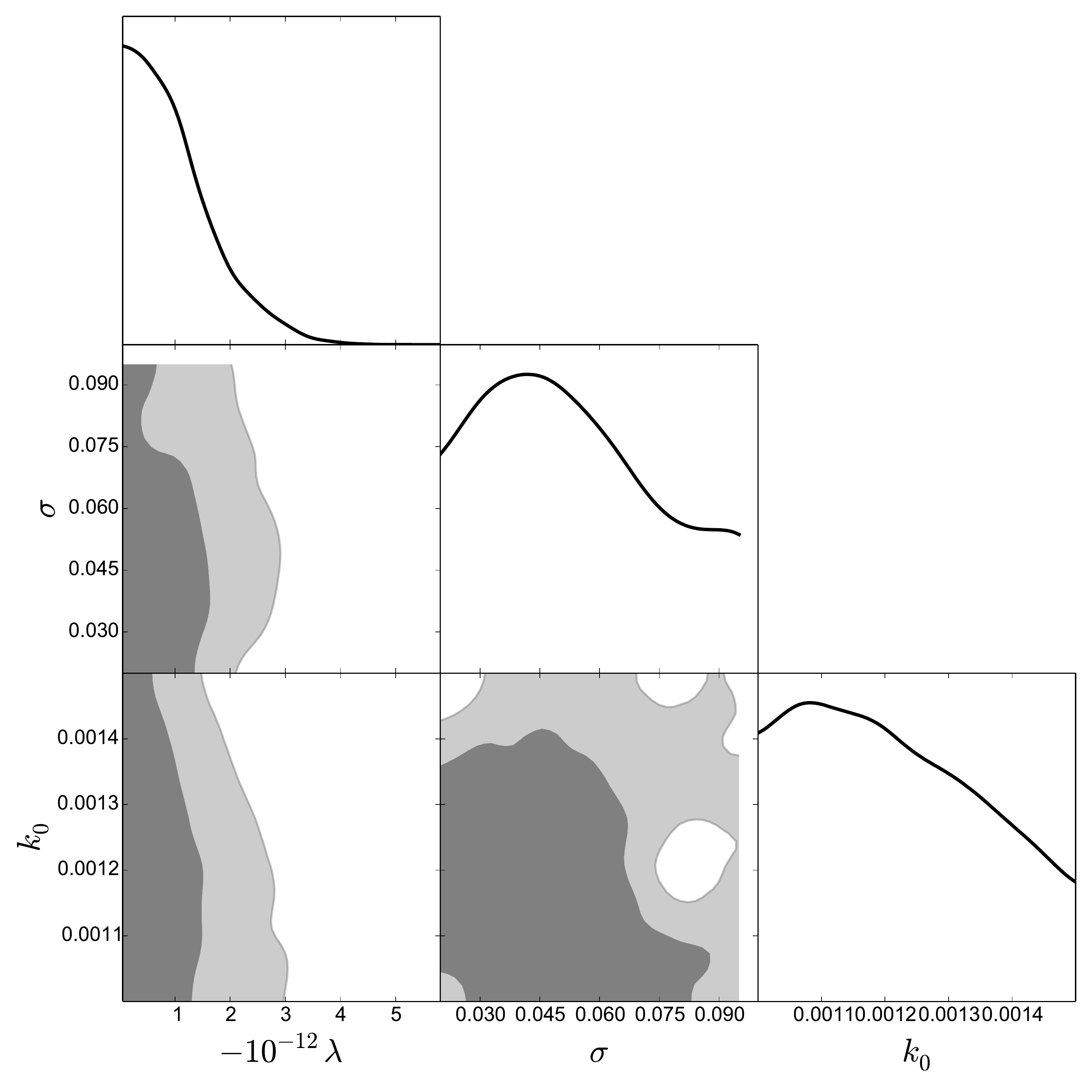}
 \caption{The results of the data fitting analysis for the parameters
 $\lambda$, $\sigma$ and $k_0$ are shown
 for the model with local features.
 }
\label{fig:infl_params}
\end{figure}

\begin{figure}
 \begin{minipage}{.45\textwidth}
  \includegraphics[scale=0.6]{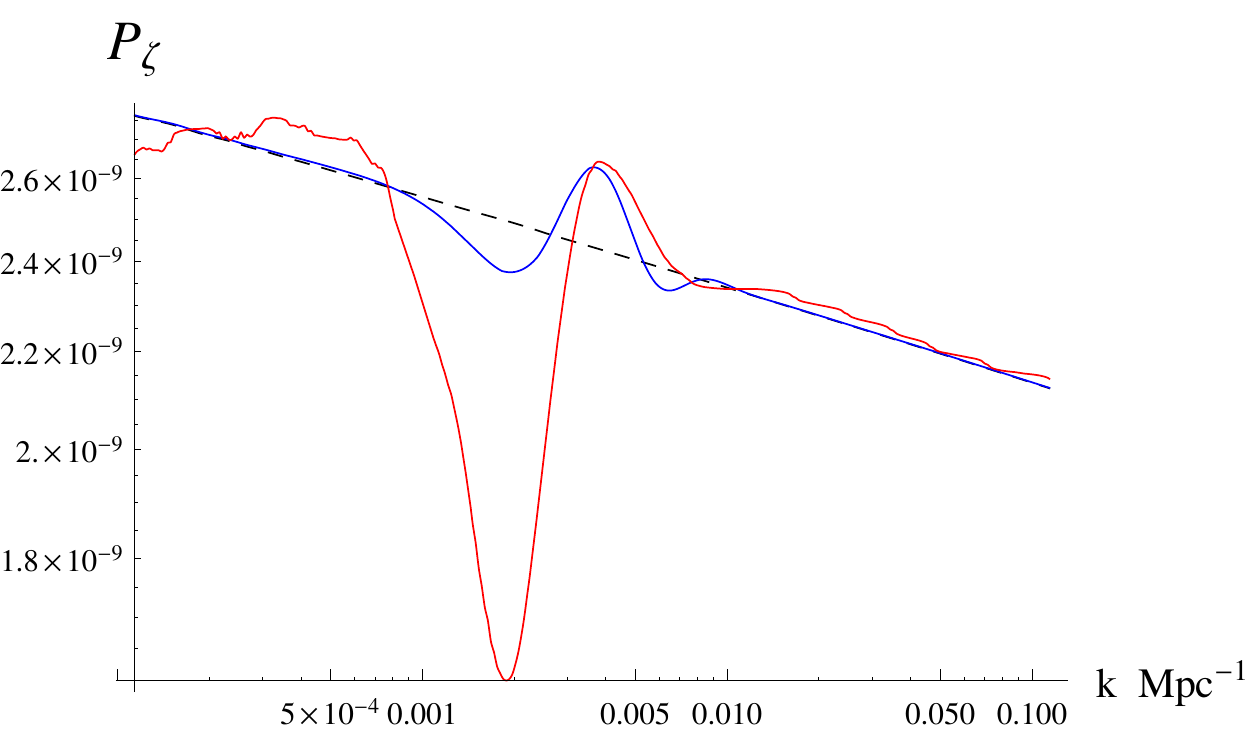}
  \end{minipage}
 \begin{minipage}{.45\textwidth}
  \includegraphics[scale=0.6]{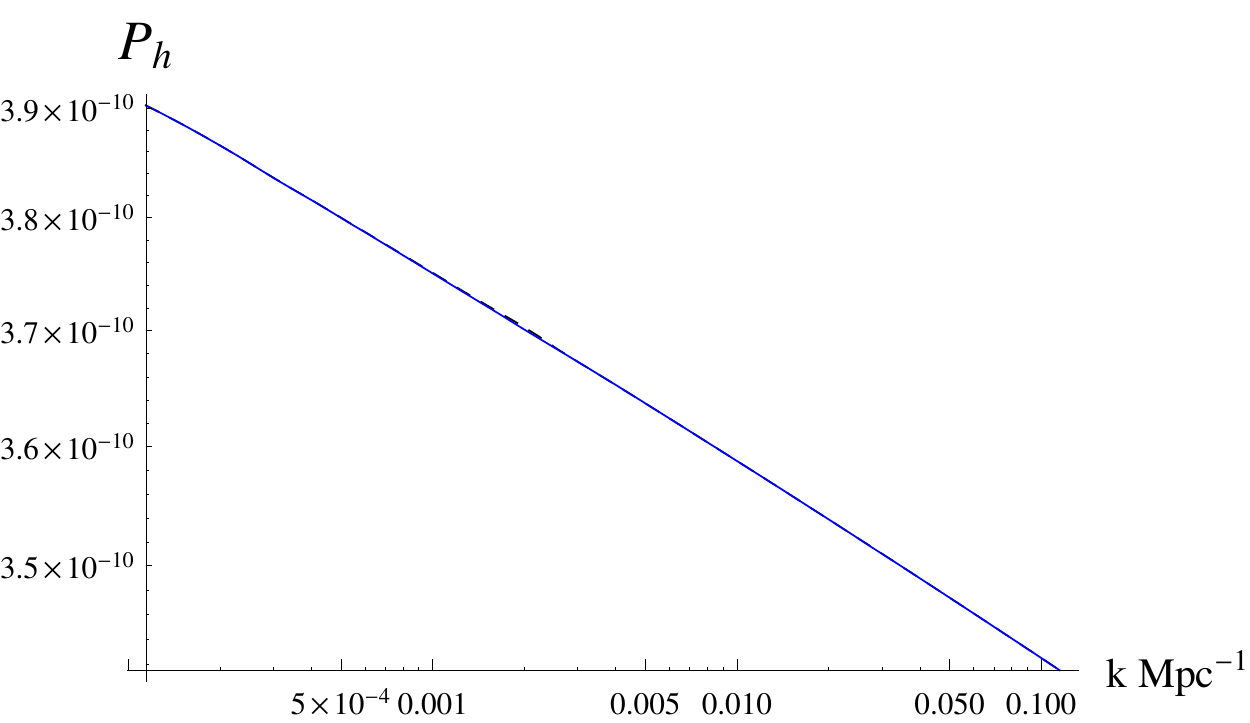}
 \end{minipage}
 \caption{The numerically computed spectrum of the primordial curvature fluctuations $P_{\zeta}$ and of the tensor perturbations $P_{h}$ are plotted for the best fit values in Tab.~\ref{tab:results}: $\lambda=-1.12\times10^{-12},  \sigma=0.053$, and 
$k_0=1.13\times 10^{-3}$ (blue). On the left, the red lines correspond to the best-fit reconstructed primordial power spectrum from Ref.~\cite{DiValentino:2016ikp}. The dashed lines correspond to the featureless spectrum.}
\label{PandPhbestfit}
\end{figure}

\begin{figure}
	\centering
  \includegraphics[width=\textwidth]{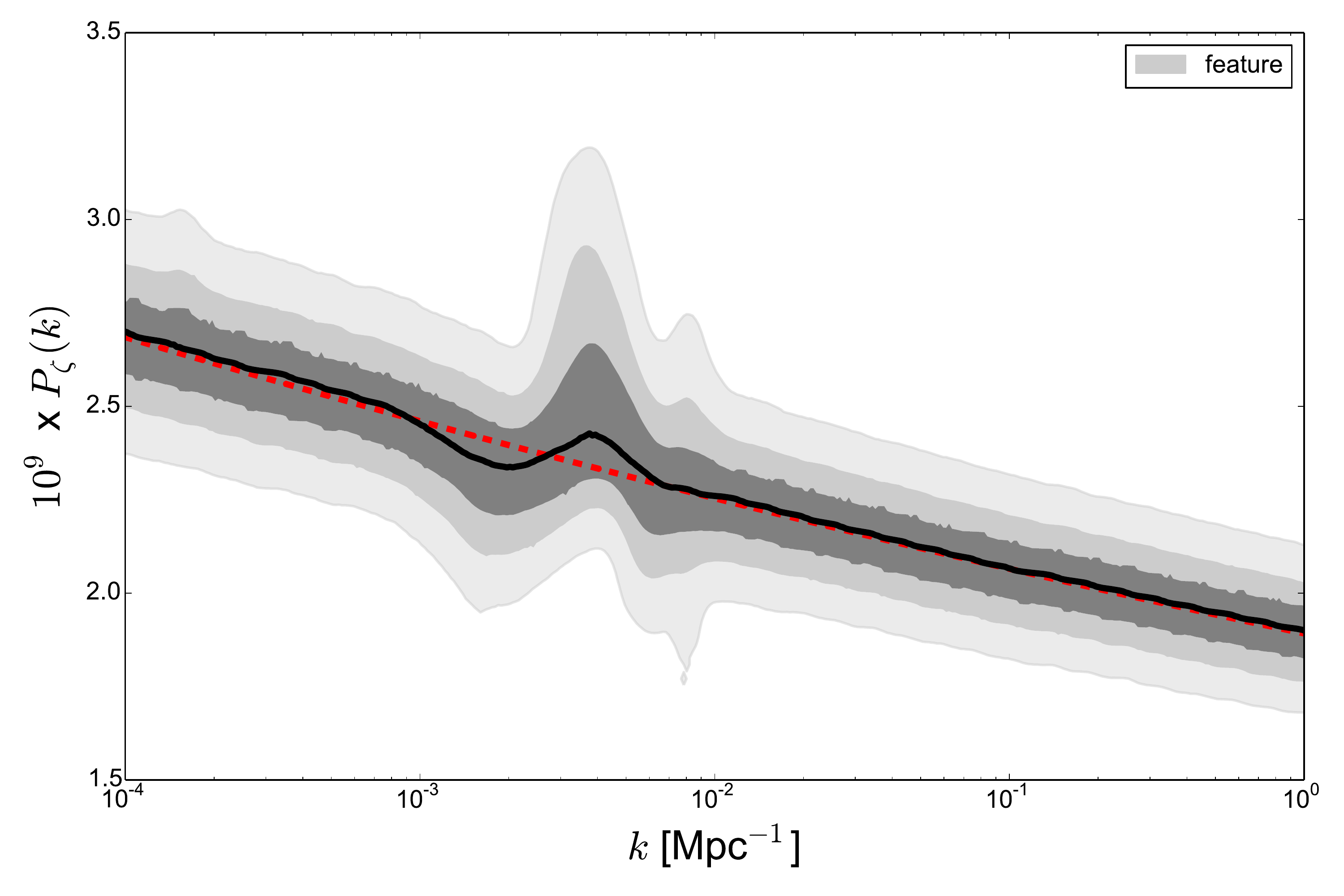}
 \caption{The 1, 2 and 3$\sigma$ constraints obtained from observational data analysis are plotted for the primordial curvature perturbations spectrum for the model with local features.  
 The  spectrum for the featureless model is plotted with a red line.
 }
\label{fig:scal_bands}
\end{figure}

% \begin{figure}
% 	\centering
%  \includegraphics[width=\textwidth]{pps_bands_tens.pdf}
%  \caption{The 1, 2 and 3$\sigma$ constraints obtained from observational data analysis are plotted for the primordial tensor perturbations power spectrum for the model with features.  
%  The  spectrum for the featureless model is plotted with a red line.}
% \label{fig:tens_bands}
% \end{figure}
%*********************************

The results of the data fitting analysis are reported in Tab.~\ref{tab:results} and
in Figs.~\ref{fig:lcdm_params} to \ref{fig:scal_bands}.

In Tab.~\ref{tab:results} we show the best-fit values written inside brackets and the 1$\sigma$ constraints
of the parameters. It should be noted that the bounds we obtain are more stringent than the Planck ones because $n_s$ is not a free parameter. Fixing the value of scalar spectral index reduces the confidence  ranges for the others parameters, and consequently our bounds are smaller. If we had left free the potential of the inflaton in a  generic monomial form $V_0 \sim \phi^n$, then we could have obtained larger bounds as in the Planck team analysis where $n_s$ is a free parameter. This could be done in a future work, but it goes beyond the scope of the present paper.

Comparing the results obtained with and without feature we can see that the presence of the LF  has no impact on
the background cosmological parameters. % that describe the evolution:the fact that we consider a feature of the local type means that the overall behavior of the primordial power spectrum does not change,
%and consequently there is no overall influence on the final CMB spectra.
This is  clear from the marginalized 1D and 2D plots in Fig.~\ref{fig:lcdm_params}.
The effect of the feature is evident around the location of the bump of the CMB temperature spectrum (see Fig.~\ref{clTT}),
and it corresponds to an improvement of the total $\chi^2$.
As reported in Tab.~\ref{tab:results}, the improvement comes from the $\chi^2$ of the low-$\ell$ Planck likelihood. %s,
% while there is a small degrading of the fit for the BKP spectrum.
Our analysis cannot be compared with the Planck results \cite{Ade:2015xua,Ade:2015tva},
we are assuming the $\phi^2$ inflationary model
instead of using a phenomenological approach with independent $n_s$ and $r$.
Quadratic inflation corresponds to high values of $r$
which are not in agreement with the Planck best-fit model obtained using $n_s$ and $r$ as independent parameters.
The effects of the feature improve the $\chi^2$ with respect to the featureless $\phi^2$ case,
but this improvement is not large enough to make it competitive with other models.
Nevertheless, the same LF could be applied to other
inflationary scenarios to produce an analogous improvement
of the $\chi^2$.
The analyses of the effects of the LF for inflationary models
that are in better agreement with the observed CMB spectra
are left for future studies.

In Fig.~\ref{fig:infl_params} we show the 1D marginalized posterior distributions and the correlations between the feature parameters.
From the correlation plot between $\lambda$ and $k_0$ we can see that
the size of the feature can be larger if the feature is located at a smaller wavemode $k_0$.
This is because the CMB temperature spectrum does not allow any wiggles above $\ell\simeq60$, thus limiting the amplitude of the feature.
The 2D plots for the parameter $\sigma$  seem to show that there is no lower bound on it.
This is not in tension with the 1$\sigma$ constraints on the $\sigma$ parameter reported in Tab.~\ref{tab:results},
because of volume effects that occur in the Bayesian marginalization procedure.
The preference for a non-minimum value of $\sigma$ is mild, indeed 
there is no lower bound at 2$\sigma$ confidence level.

The constraints on the primordial scalar spectrum are shown in Figs.~\ref{PandPhbestfit} and \ref{fig:scal_bands}.
In the left panel of Fig.~\ref{PandPhbestfit} we compare the best-fit primordial power spectrum of scalar perturbations obtained in our analysis (blue) and the reconstructed one from Ref.~\cite{DiValentino:2016ikp}.
The comparison underlines how a local feature can reproduce the behaviour of the primordial spectrum,
but further studies, which will be presented in some future work,
on the feature potential are required in order to obtain a perfect agreement.
The right panel of the same Fig.~\ref{PandPhbestfit} shows that the effect of the feature is very small in the tensor spectrum.
In Fig.~\ref{fig:scal_bands} we plot the marginalized constraints on the primordial scalar spectrum.
The 1, 2, and 3$\sigma$ bands refer to the model with LF,
while the solid black line shows the corresponding best-fit spectrum,
computed from the entire set of primordial spectra obtained from the MCMC scan.
The red dashed line shows the spectrum obtained for the same cosmological parameters but without the feature.
% In Fig.~\ref{fig:tens_bands} we plot the same as in Fig.~\ref{fig:scal_bands}, but for the tensor spectrum.
From the figures we can note that the effects that the LF brings are more important for the scalar spectrum,
while they are negligible for the tensor spectrum.
For this reason, we do not show the same plot as for Fig.~\ref{fig:scal_bands} for the tensor spectrum.
%In both cases the featureless case lies inside the 1$\sigma$ confidence band.

%As it is already known, the $\phi^2$ inflationary model is not very good
%in explaining the most recent CMB data
%and the global $\chi^2$ of the \lcdm\ fit \cite{Ade:2015tva} is much lower.
%The presence of an high amount of tensor modes in the $\phi^2$ model, indeed,
%is disfavored in particular by the most recent measurements on the tensor spectrum.
%What we want to show, however, is simply that the presence of the feature,
%assuming no variations in the cosmological parameters,
%allows to slightly improve the fit of the CMB spectrum.
%Further studies of the effects of local features on different  inflation potentials
%are left for future analysis.

\section{Conclusions}
We have studied the effects of local features in the inflaton potential on the spectra of primordial curvature perturbations and their impact on the temperature anisotropies of the CMB. 
In order to study the effects on the CMB spectrum we have modified the \texttt{CAMB} and \texttt{CosmoMC} codes
in order to use a non-standard power-law power spectrum for the primordial perturbations,
to take into account the presence of the local feature.
We have performed a best fit analysis of CMB temperature and polarization data from %Bicep2/Keck and
Planck.
%In Tab. \ref{tab:results} and Figs. \ref{fig:lcdm_params} to \ref{fig:tens_bands} we have reported the results of the analyses. 
We have found no significant effects on cosmological parameters related to the propagation of CMB photons after decoupling, 
while  LF improve the fit of the CMB temperature and polarization data. 
% as can be seen from the $\chi^2$ of Tab. \ref{tab:results}. 
We have also confirmed the theoretical expectation that local features do not affect the primordial power spectrum at scales far from the characteristic scale $k_0$, which leaves the horizon around the feature time.

In the future it will be interesting to analyze the effects of local features in order to explain other deviations of the CMB spectrum, such as for example the anomalies occurring around $l\approx 800$ \cite{Hazra:2014jwa}. It will also be important to study  the effects of LF in inflationary models with different featureless $V_0$ potentials, and to compare them to the effects of branch features.
%as can be seen in the marginalized 1D and 2D plots in Fig.~\ref{fig:lcdm_params} and, consequently there is no overall influence on the final CMB spectra. 

%In Figs.~\ref{fig:scal_bands} and \ref{fig:tens_bands} we show the constraints on the primordial scalar and tensor spectra, respectively. 
%The best-fit spectrum is obtained from the marginalization of the whole set of spectra obtained from the MCMC scan.

%It is left for future studies to consider a different inflationary model $V_0(\phi)$ besides the $\phi^2$ used in this work since it is not in very good agreement with the most recent CMB data.
\acknowledgments
This work was supported by the European Union (European Social Fund, ESF) and Greek national funds under the “ARISTEIA II” Action.
%, the Dedicacion exclusica and Sostenibilidad programs at UDEA, the UDEA CODI
%project IN10219CE.
Part of the work of S.G. was supported by the Theoretical Astroparticle Physics research Grant No. 2012CPPYP7 under the Program PRIN 2012 funded by the Ministero dell'Istruzione, Universit\`a e della Ricerca (MIUR), and in part is also supported by the Spanish grants FPA2014-58183-P, Multidark CSD2009-00064 and SEV-2014-0398
(MINECO), and PROMETEOII/2014/084 (Generalitat Valenciana). The work of A.G.C. was supported by the Colombian Department of Science, Technology, and Innovation COLCIENCIAS research Grant No. 617-2013. A.G.C. acknowledges the partial support from the International Center for Relativistic Astrophysics Network ICRANet. For part of the calculations we used the Cloud infrastructure of the Centro di Calcolo in the Torino section of INFN.
AER work was supported by the Dedicacion exclusica and Sostenibilidad programs at
UDEA, the UDEA CODI project 2015-4044 and 2016-10945, and Colciencias mobility program.

% \bibliography{Bibliography}
% \bibliographystyle{h-physrev4}
% \input{main.bbl}
% \end{document}

\end{document}